# Josephson diode effect from Cooper pair momentum in a topological semimetal

Banabir Pal[1]*, Anirban Chakraborty[1]*, Pranava K. Sivakumar[1]*, Margarita Davydova[2]*, Ajesh K. Gopi[1], Avanindra K. Pandeya[1], Jonas A. Krieger[1], Yang Zhang[2], Mihir Date[1], Sailong Ju[3], Noah Yuan[2], Niels B.M. Schröter[1], Liang Fu[2] and Stuart S.P. Parkin[1]

1. Max Planck Institute of Microstructure Physics, Halle (Saale) 06120, Germany
2. Department of Physics, Massachusetts Institute of Technology, Cambridge, MA 02139, USA
3. Swiss Light Source, Paul Scherrer Institute, CH-5232 Villigen PSI, Switzerland
* Equal author contributions

**In the presence of an external magnetic field Cooper pairs in noncentrosymmetric superconductors can acquire finite momentum. Recent theory predicts that such finite-momentum pairing can lead to an asymmetric critical current, where a dissipationless supercurrent can flow along one direction but not the opposite. However, to date this has not been observed. Here we report the discovery of a giant Josephson diode effect (JDE) in Josephson junctions formed from a type II Dirac semimetal, $NiTe_2$. A distinguishing feature is that the asymmetry in the critical current depends sensitively on the magnitude and direction of an applied magnetic field and achieves its maximum value of ~60% when the magnetic field is perpendicular to the current and is of the order of just 10 mT. Moreover the asymmetry changes sign several times with increasing field. These characteristic features are accounted for in a theoretical model based on finite-momentum Cooper pairing derived from spin-helical topological surface states, in an otherwise centrosymmetric system. The finite pairing momentum is further established, and its value determined, from the evolution of the interference pattern under an in-plane magnetic field. The observed giant magnitude of the asymmetry in critical current and the clear exposition of its underlying mechanism paves the way to building novel superconducting computing devices using the Josephson diode effect.**

Semiconductor junctions, which exhibit direction-dependent non-reciprocal responses, are essential to modernday electronics [1-3]. On the other hand, a key component of many quantum technologies is the superconducting Josephson junction (JJ) where two superconductors are coupled via a weak link [4]. JJs can be used for quantum sensing of small magnetic fields [5,6], single-photon detection [7-9] and quantum computation [10-12]. Despite the longstanding research on superconductivity, the realization of the superconducting analogue of the diode effect, i.e., the dissipationless flow of supercurrent along one direction but not the other, has been reported only recently in superconducting thin films[13] and Josephson junctions[14,15]. However, a clear experimental evidence for a specific mechanism leading to this effect is lacking. Recent theoretical work[16-18] has proposed that in 2D superconductors with strong spin-orbit coupling under an in-plane magnetic field, Cooper pairs can acquire a finite momentum and give rise to a diode effect, where the direction of the Cooper pair momentum determines the polarity of the effect. At the same time, a theoretical description for a field induced diode effect in Josephson junctions (the Josephson diode effect) has not yet been formulated.

In this work, we report the discovery of a giant Josephson diode effect (JDE) in a JJ in which a type-II Dirac semimetal $NiTe_2$ couples two superconducting electrodes and provide clear evidence of its interrelation with the presence of finite momentum Cooper pairing. The effect depends sensitively on the presence on a small in-plane magnetic field. The non-reciprocity, $\Delta I_c$, i.e. the difference between the critical currents for opposite current directions, is antisymmetric under an applied in-plane magnetic field $B_{ip}$. $\Delta I_c$ depends strongly on the angle between $B_{ip}$ and the current direction and is largest when $B_{ip}$ is perpendicular to the current and vanishes when $B_{ip}$ is parallel to it. Moreover, we also observe multiple sign reversals in $\Delta I_c$ when the magnitude of $B_{ip}$ is varied. Our phenomenological theory shows that the presence of a finite Cooper pair momentum (FCPM) and a non-sinusoidal current-phase relation account for all the salient features of the observed JDE, including the angular, temperature and magnetic field dependences of the observed $\Delta I_c$. The distinct evolution of the interference pattern under the in-plane magnetic field further establishes the presence of finite Cooper pair momentum in this system. Here, the finite momentum pairing results from the momentum shift of topological surface states of $NiTe_2$ under an in-plane magnetic field, a feature enabled by spin-momentum locking. By ARPES measurements and comparison with DFT calculations, we further identify a surface pocket with a very small Fermi energy that is ideal for inducing a large Cooper pair momentum and, therefore, a large Josephson diode effect. This paper presents the temperature, field and angle dependence of the Josephson diode effect, and is the only work to date in which the fundamental properties of the effect can be explained within a single model providing clear evidence that the features are consistent with finite momentum of the Cooper pairs in the presence of magnetic field.

NiTe$_2$ crystallizes in a CdI$_2$ type trigonal crystal structure with the space group $P\bar{3}m1$, which is centrosymmetric [19,20] (see supplementary information (SI) for details). This two-dimensional van der Waals material is a type-II Dirac semimetal that hosts several spin-helical topological surface states [19,20]. As we shall discuss later, these surface states play a key role in the JDE. JJ devices were fabricated on NiTe$_2$ flakes that were first mechanically exfoliated from a single crystal (see SI for details). Fig. 1a shows optical images of several JJ devices formed on a single NiTe$_2$ flake, where the edge-to-edge separation ($d$) between the superconducting contacts (formed from 2 nm Ti/ 30 nm Nb/ 20 nm Au) in each device is different. A schematic of the JJ device is shown in Fig. 1b-c, in the absence and presence of Josephson current, respectively, where $x$ is parallel to the current direction and $z$ is the out-of-plane direction. The temperature dependence of the resistance of the device with $d = 350$ nm (Fig. 1d) shows two transitions: the first transition ($T_{SC}$) at ~ 5.3 K is related to the superconducting electrodes [21]. A second transition ($T_J$) takes place at a lower temperature when the device enters the Josephson transport regime such that a supercurrent flows through the NiTe$_2$ layer. The dependence of both $T_{SC}$ and $T_J$ as a function of the edge-to-edge separation, $d$, between the electrodes is shown in the inset of Fig. 1d. While $T_{SC}$ is independent of $d$, $T_J$ decreases with increasing $d$, which corroborates that $T_J$ corresponds to the superconducting proximity transition of the JJ device [22].

To observe the JDE (see illustration in Fig. 1b-c), we carried out current versus voltage (*I-V*) measurements as a function of temperature and magnetic field. Fig. 1e shows *I-V* curves in the presence of an in-plane magnetic field $B_y \sim 20$ mT perpendicular to the direction of the current for the device with $d = 350$ nm. The device exhibits four different values of the critical current with a large hysteresis indicating that the JJs are in the underdamped regime [23]. During the negative to positive current sweep (from -50 μA to +50 μA) the device shows two critical currents $I_{r-}$ and $I_{c+}$ whereas during a positive to negative current sweep (from +50 μA to -50 μA) the device exhibits two other critical currents $I_{r+}$ and $I_{c-}$. In the rest of the paper, we concern ourselves with the behavior of the critical currents $I_{c-}$ and $I_{c+}$, which correspond to the critical values of the supercurrent when the system is still superconducting. For small magnetic fields, we find that the absolute magnitude of $I_{c-}$ is clearly much larger than that of $I_{c+}$ (Fig. 1e) (see SI for zero field data where $I_{c+} = |I_{c-}|$). These different values of $I_{c+}$ and $|I_{c-}|$ mean that when the absolute value of the applied current lies between $I_{c+}$ and $|I_{c-}|$, the system will behave as a superconductor for the current along one direction while a normal dissipative metal for the current along the opposite direction. We use this difference to demonstrate a clear rectification effect, as shown in Fig. 1f, that occurs for currents which are larger than $I_{c+}$ but smaller than $|I_{c-}|$.

To probe the origin of the JDE, the evolution of $\Delta I_c$ ($\Delta I_c \equiv I_{c+} - |I_{c-}|$) was examined as a function of applied in-plane magnetic field at various temperatures and angles with respect to the current direction. The dependence of $\Delta I_c$ on $B_y$ (field parallel to the y axis and perpendicular to the current) at different temperatures demonstrates that $\Delta I_c$ is antisymmetric with respect to $B_y$ (Fig. 2a). At 60 mK, $\Delta I_c$ exhibits a maximum and a minimum value at $B_y = \mp 12$ mT, respectively (Fig. 2a) and the ratio $\frac{\Delta I_c}{\langle I_c \rangle}$ is as large as 60%, where $\langle I_c \rangle = \frac{(I_{c+} + |I_{c-}|)}{2}$. Such a large magnitude of $\frac{\Delta I_c}{\langle I_c \rangle}$ at a low magnetic field (~ 12 mT) makes this system unique, as compared to previous reports where either the magnitude of $\frac{\Delta I_c}{\langle I_c \rangle}$ was found to be small, or a large magnetic field was required to observe a significant effect [13-15]. We also observe multiple sign reversals in $\Delta I_c$ when $B_{ip}$ is increased (see SI for details), a previously unobserved but interesting dependence that is critical to unraveling the origin of the JDE, as we discuss below.

The dependence of $\Delta I_c$ on the direction of the in-plane magnetic field with respect to the current is shown in Fig. 2b-c, for several field strengths. At small fields, $|\Delta I_c|$ is largest when the field is perpendicular to the current ($\theta = 0°/\pm 180°$, where $\theta$ is the in-plane angle measured with respect to the *y*-axis) and vanishes when the field and current are parallel ($\theta = \pm 90°$). With regard to the temperature dependence of $\Delta I_c$, we find that the magnitude of $\Delta I_c$ increases monotonically as the temperature is lowered (Fig. 2a). For a quantitative understanding, the temperature dependence of $\Delta I_c$ for $B_y = 12$ mT (the field at which $\Delta I_c$ takes the largest value) is shown in Fig. 2d. At temperatures near $T_J$, the variation of $\Delta I_c$ with temperature can be well fitted by the equation $\Delta I_c = \alpha (T - T_J)^2$ (inset of Fig. 2d).

We propose a possible origin of the JDE as follows. At temperatures close to $T_J$ at which superconducting correlations develop in the proximitized region [22,24,25] and the Josephson effect emerges (Fig. 1d), the free energy $F$ of our system can be expanded in powers of the superconducting order parameters of the two superconducting electrodes, $\Delta_{1,2}$ in the proximitized regions:

$$F = F_0 - \gamma_1 \Delta_1^* \Delta_2 - \frac{1}{2}\gamma_2 (\Delta_1^* \Delta_2)^2 + c.c. + \cdots \qquad (1)$$

where $F_0$ is the free energy in the absence of Josephson coupling, and $\gamma_1$ and $\gamma_2$ denote, respectively, first- and second-order Cooper pair tunneling processes across the weak link. The presence of higher harmonics account for a non-sinusoidal current-phase relation, as commonly observed in superconductor-normal metal-superconductor (SNS) junctions with high transmission. Importantly, in the absence of time-reversal and inversion symmetries, $\gamma_{1,2}$ are complex numbers, which makes the critical current non-reciprocal, as we show below.

Expressing the order parameters $\Delta_{1,2}$ in terms of their amplitude and phase as $\Delta_{1,2}= \Delta e^{i\varphi_{1,2}}$, the free energy takes the form $F = F_0 - 2|\gamma_1|\Delta^2 \cos\varphi - |\gamma_2|\Delta^4 \cos(2\varphi + \delta)$, where $\varphi = \varphi_2 - \varphi_1 + arg(\gamma_1)$ is effectively the phase difference between the two superconducting regions. Note that, indeed, when both time-reversal and inversion symmetries are broken, the phase-shifted Josephson junction is realized, as was observed in Ref. 26. $\delta = arg(\gamma_2) - 2\,arg(\gamma_1)$ is the phase shift associated with the interference between the first- ($\gamma_1$) and second-order ($\gamma_2$) Cooper pair tunneling processes. The Josephson current-phase relation then includes the second harmonic:

$$I(\phi) = \frac{2\pi}{\Phi_0}\frac{\partial F}{\partial \varphi} = \frac{4e}{\hbar}\{\Delta^2|\gamma_1|\sin\varphi + \Delta^4|\gamma_2|\sin(2\varphi + \delta)\} \quad (2)$$

where $\Phi_0 = \frac{h}{2e}$ is the superconducting flux quantum. When $\Delta^4|\gamma_2|$ is small, the critical current of the Josephson junction is reached near a phase difference $\varphi \approx \pm\pi/2$, and equals:

$$I_{c\pm} \approx |I(\pm\frac{\pi}{2})| = \frac{4e}{\hbar}\{\Delta^2|\gamma_1| \mp \Delta^4|\gamma_2|\sin\delta\} \quad (3)$$

The non-reciprocal part of the critical current is proportional to $\Delta^4$:

$$\Delta I_c \equiv I_{c+} - |I_{c-}| = -\frac{8e}{\hbar}\Delta^4|\gamma_2|\sin\delta \quad (4)$$

Since the pairing potential in the proximitized layer behaves as $\Delta \propto \sqrt{1-\frac{T}{T_J}}$ [22,24,25], the temperature dependence of $\Delta I_c$ near $T_J$ is then given by:

$$\Delta I_c \propto \Delta^4 \propto (T - T_J)^2 \quad (5)$$

which explains our experimentally measured temperature dependence of $\Delta I_c$ that is shown in the inset of Fig. 2d.

Another feature is that $\Delta I_c$ can change sign as the applied field increases (see, e.g. near $B_{ip} \sim 22$ mT in Fig. 2a and 2c). Such a sign reversal in $\Delta I_c$ can be reproduced by including the field dependence of the order parameters $\Delta_{1,2} \propto \sqrt{1-\left(\frac{|B|}{B_c}\right)^2}$ (where $B_c$ is the critical field in the proximitized region) and of the phase shift $\delta$ due to the Cooper pair momentum. The in-plane magnetic field $B_{ip}$ can induce a finite Cooper pair momentum via the screening current [27], and/or through the Zeeman effect on topological surface states [28-30]. Indeed, due to spin-momentum locking, the spin-helical topological surface states of NiTe$_2$ (discussed in more detail below) acquire a momentum shift $q_x$ under $B_y$, which effectively turns the proximitized region into a finite-momentum superconductor (see Fig. 1b-c). The presence of Cooper pair momentum results in a phase shift accumulated during the Cooper pair propagation across the junction: $\delta \approx 2q_x d$. At small values of the field, $q_x$ must be linear in $B_y$, so that:

$$\delta \approx \pi \frac{B_y}{B_d} \tag{6}$$

where $B_d$ is a property of the junction geometry and material that can, in principle, be determined based on the specific microscopic origin of the field-induced Cooper pair momentum. As a result, $\Delta I_c$ will have the following field dependence:

$$\Delta I_c \propto \Delta^4 \sin \delta \propto \left[1 - \left(\frac{|B|}{B_c}\right)^2\right]^2 \sin\left(\pi \frac{B_y}{B_d}\right) \tag{7}$$

Depending on the ratio $\frac{B_d}{B_c}$, different scenarios can be realized from this equation, and for $\frac{1}{2(n+1)} < \frac{B_d}{B_c} < \frac{1}{n}$, there are $n$ sign reversals in $\Delta I_c$ when the magnetic field is applied in y-direction.

Figs. 2e-f show the dependencies of $\Delta I_c$ on the magnitude of $B_y$ and the direction of the in-plane magnetic field as obtained from our phenomenological model (eq. (7)), where we used $B_c = 45$ mT and $B_d \approx 22$ mT (see SI for details). In order to explain the angular direction dependence, we take into account that the non-reciprocal part of the current is proportional only to the x-component of the momentum shift ($q_x$) that is proportional to the $B_y (= B_{ip} \cos\theta)$ component of the in-plane magnetic field. In Fig. 2f we see that in each domain $-\frac{\pi}{2} + \pi n < \theta < \frac{\pi}{2} + \pi n$, the sign reversal of $\Delta I_c$ occurs where the condition $\sin\left(\pi \frac{B_{ip} \cos\theta}{B_d}\right) = 0$ is fulfilled, which one can see clearly in Fig. 2c. Thus, our model successfully captures the main features of the JDE as seen in our experimental data.

To confirm the emergence of a finite Cooper pair momentum under an in-plane magnetic field, we examine the evolution of the interference pattern ($\frac{dV}{dI}$ vs $B_z$) with the field $B_x$ parallel to the current direction (Fig. 3b). This interference pattern has a similar resemblance with the Fraunhofer pattern, such that, a higher $I_c$ in the latter translates to a lower $\frac{dV}{dI}$ in the former [29,30]. For this in-plane field orientation ($B_x$), the Cooper pairs acquire a finite momentum $2q_y$ along the y-direction (Fig. 3a), which should not generate a JDE but is expected to change the interference pattern. When a Cooper pair tunnels from position ($x = 0, y_1$) in the left superconductor with order parameter $\Delta_1(y_1) = \Delta\, e^{2iq_y y_1}$ to the superconductor on the right at ($x = d_{eff}, y_2$) with $\Delta_2(y_2) = \Delta\, e^{2iq_y y_2 + i\varphi_0}$ (where $d_{eff} = d + 2\lambda$ is the effective length of the junction with $\lambda = 140$ nm the London penetration depth in Nb [31]), the contribution of this trajectory to the current involves a phase factor proportional to the Cooper pair momentum [29,30]:

$$\Delta\varphi = 2q_y(y_2 - y_1) \tag{8}$$

in addition to the usual phase factor $\frac{2\pi B_z d_{eff}(y_1+y_2)}{\Phi_0}$ due to the magnetic flux associated with $B_z$. The interference pattern is a result of interference from all such trajectories and is shown in Figs. 3b-c.

Due to the additional phase $\Delta\varphi$ from the in-plane field induced Cooper pair momentum, the interference pattern evolves as the in-plane field is increased, splitting into two branches. The Cooper pair momentum $2q_y$ can be extracted from the slope of the side branches [29,30], as indicated by the solid lines in Fig. 3b. For the $d = 350$ nm device, we estimate the average slope to be $\frac{B_x}{B_z} \approx 13$ (see SI for more details). In Fig. 3c we show the calculated critical Josephson current, obtained by summing over quasi-classical trajectories (see SI for details), which has a qualitatively similar behavior to the differential resistance in Fig. 3b, with the same period of oscillations (~ 0.8 mT) and the slope of the side branches. The slope of the side branches can be expressed as:

$$\frac{2q_y}{B_z} \approx \frac{\pi d_{eff}}{\Phi_0} \tag{9}$$

From this and the value of the slope $\frac{B_x}{B_z}$ extracted from the experiment, we find that at $B_x = 12$ mT, the Cooper pair momentum is $2q_y \approx 1.6 \times 10^6$ m$^{-1}$.

Let us compare the estimate of the Cooper pairing momentum based on the evolution of the interference pattern with the results of the JDE measurements above. The maximum JDE is achieved when the phase shift $\delta = 2q_x d$ equals approximately $0.5\pi$, which corresponds to the Cooper pair momentum $2q_x = \frac{\delta}{d} \approx 4.5 \times 10^6$ m$^{-1}$ at $B_x = 12$ mT. While the JDE and interference peak splitting are measured under in-plane magnetic fields along two orthogonal directions, the field-induced Cooper pair momenta $2q_x$ and $2q_y$ estimated from these measurements are of the same order of magnitude, further strengthening our conclusion that both effects arise from the finite momentum of the Cooper pairs.

It is remarkable that a JJ device comprising a centrosymmetric material such as NiTe$_2$ exhibits a large JDE effect since Cooper pairs with finite momentum require a broken inversion symmetry. Therefore, the origin of the JDE is likely to be related to the surface electronic structure of NiTe$_2$ where the inversion symmetry is naturally broken. Previous studies [19,20] have reported the presence of spin-polarized surface states in NiTe$_2$ that cross the Fermi-level. Here we have used angle-resolved photoelectron spectroscopy (ARPES) to examine these surface states and estimate their Fermi energy and velocity (see SI). The energy-momentum dispersion of two topological surface states (SS) along the $\bar{\Gamma} - \bar{M}$ direction are shown in Fig. 4a, which is qualitatively reproduced by our ab-initio calculations in Fig. 4b that also indicate their spin polarization. Note that the calculation slightly underestimates the energy at which the lower lying surface state merges with the valence band bulk continuum in the experiment. The origin of these surface states is a band inversion of the valence and conduction bands above the Fermi-level [20,32], which leads to the formation of a spin-helical Dirac surface state that connects the conduction and valence bands (schematic in Fig. 4c), similar to the surface state in a topological insulator. The upper branch of this Dirac surface state forms an electron pocket with a relatively small Fermi energy (~15 meV) and Fermi velocity $v_F \sim 0.4 \times 10^5$ ms$^{-1}$, which is likely to be advantageous for generating a large JDE, as we discuss below. If we assume that the displacement of the Fermi surface of the spin-polarized SS is due to the Zeeman effect, we can roughly estimate the value of the Zeeman energy based on our earlier estimation of the value of the Cooper pair momentum and the ARPES results. One finds that the in-plane magnetic field shifts the Fermi surface [33] by $q_x = \frac{g\mu_B B_y}{\hbar v_F}$, indicating that a small Fermi velocity will lead to a large shift. If we assume a characteristic Fermi velocity of $v_F \sim 0.4 \times 10^5 \, ms^{-1}$, as measured with ARPES, and using the value of $q_x$ that we estimated from the evolution of the interference map, at $B_y = 12 \, mT$ we find the Zeeman energy to be $E_z = g\mu_B B_y = \hbar v_F q_x \approx 0.03$ meV (which corresponds to a $g$-factor of ~40). The JDE is further enhanced by the large length of the junction: because the phase shift $\delta$ that is responsible for the nonreciprocity is equal to $2q_x d$, even though the Fermi surface shift $q_x$ is small in comparison to the Fermi momentum, the phase difference acquired over the junction length $d$ can be quite large.

In summary, we have shown that a JJ device involving a type-II Dirac semimetal NiTe$_2$ exhibits a large non-reciprocal critical current. The effect is antisymmetric with respect to the magnetic field perpendicular to the current direction and exhibits a $(T - T_J)^2$ temperature dependence. The evolution of the interference pattern under the application of an in-plane magnetic field provides compelling evidence for finite-momentum Cooper pairing. These experimental observations of non-reciprocal behavior induced by finite-momentum Cooper pairs and a field-induced sign reversal of non-reciprocity open new directions for future research in the field of superconductivity.

**Acknowledgments:** We thank Prof. Takis Kontos and Dr. See-Hun Yang for valuable discussions. SSPP acknowledges the Deutsche Forschungsgemeinschaft (DFG, German Research Foundation) – project no. 443406107, Priority Programme (SPP) 2244. The work at Massachusetts Institute of Technology was supported by a Simons Investigator Award from the Simons Foundation. LF was partly supported by the David and Lucile Packard foundation. J.A.K. acknowledges support by the Swiss National Science Foundation (SNF-Grant No. P500PT_203159).

**Figures and Captions**

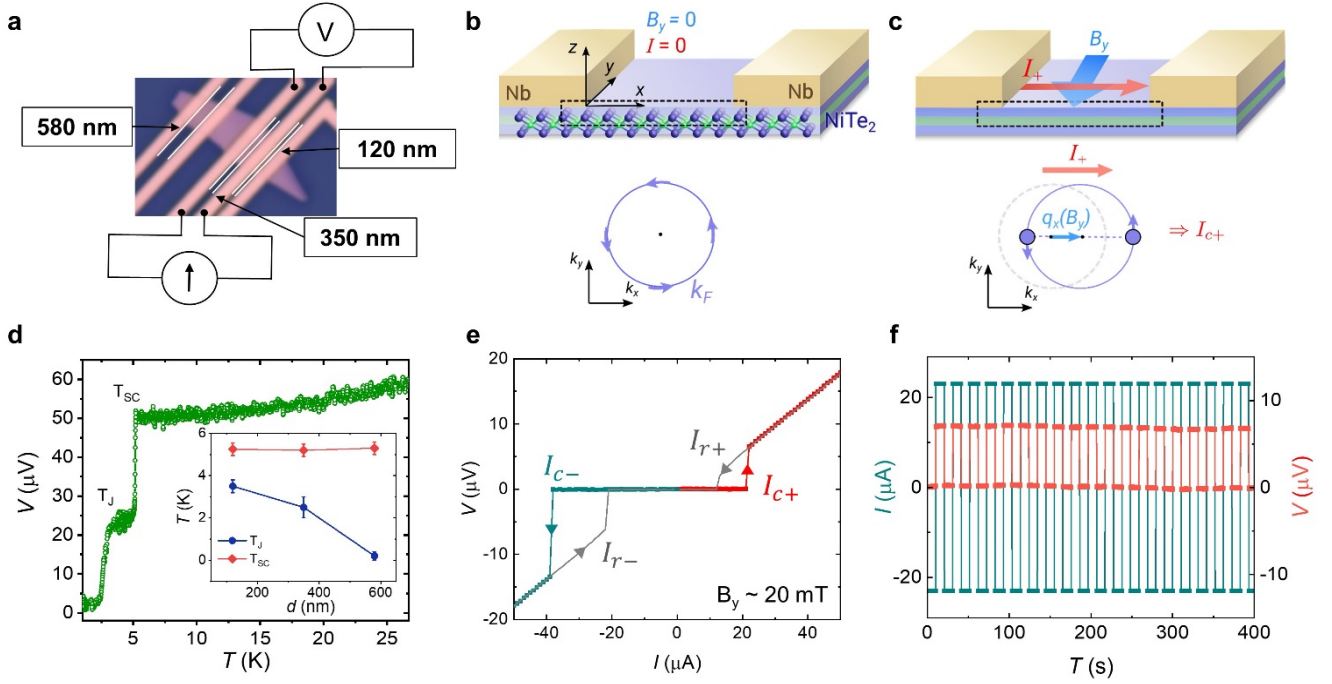

**Fig. 1| NiTe$_2$ JJ device and the observation of a Josephson Diode effect**: **a,** Optical microscopy image of several JJ devices formed on a single NiTe$_2$ exfoliated flake. The edge-to-edge spacing $d$ between the superconducting electrodes varies for each device, as shown in the figure. **b-c,** Schematic illustrations of the Josephson device and the corresponding fermi surface. In the absence of both in-plane magnetic field and the Josephson current, the weak link contains spin-orbit coupled helical surface states, as schematically shown in the bottom panel of Fig. 1b. As we discuss in the text, in the presence of an in-plane magnetic field, the helical surface state acquires a momentum shift $q_x = q_x(B_y)$, as shown in Fig. 1c. When the Josephson current flows through the junction ($I_+$ corresponds to the Josephson current flowing in the $+x$ direction), the critical current is different depending on its direction. **d,** Voltage as a function of temperature at zero field for a JJ device with $d$ = 350 nm that shows two transitions at $T_{SC}$ and $T_J$. Inset of Fig. 1d shows the variation of $T_{SC}$ and $T_J$ with $d$. **e,** I-V curve of a JJ device with $d$ = 350 nm at 20 mK and an in-plane magnetic field $B_y$ = 20 mT showing a large non-reciprocal critical current, $I_{c-}$ and $I_{c+}$. **f,** Rectification effect observed using currents between $|I_{c-}|$ and $I_{c+}$ for the same JJ device at 20 mK and $B_y$ = 20 mT.

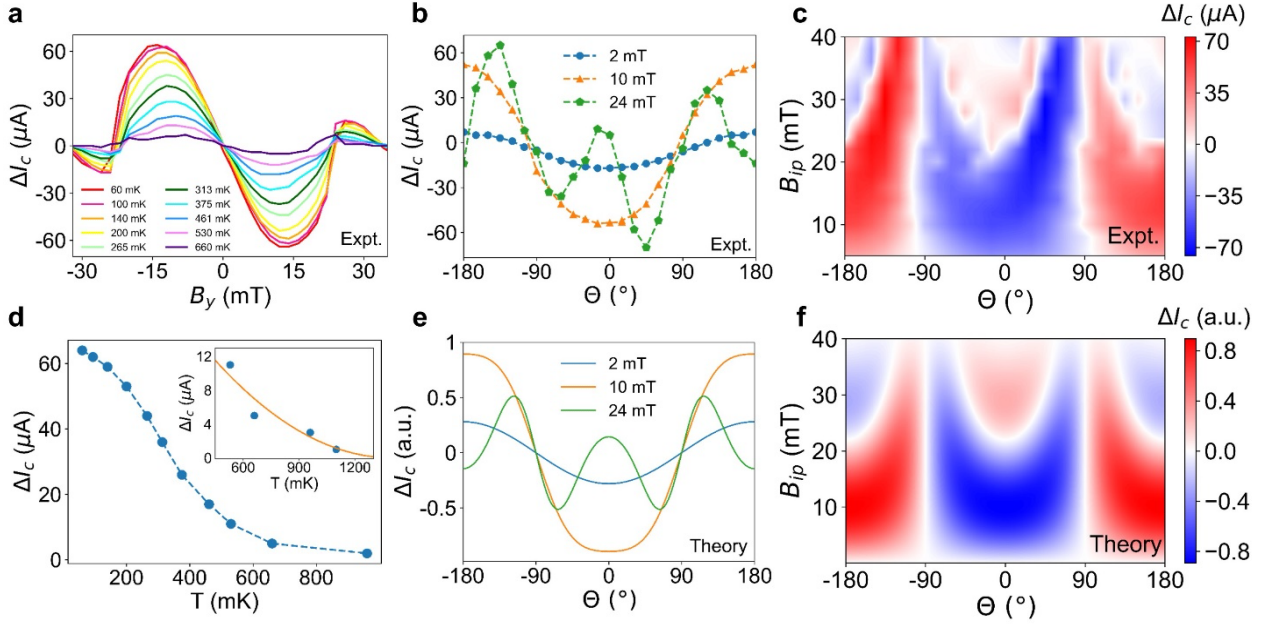

**Fig. 2| Dependence of ΔI$_c$ on in-plane magnetic field, angle and temperature. a,** Variation of ΔI$_c$ as a function of $B_y$ at selected temperatures in a JJ device with $d$ = 350 nm. **b-c,** Dependence of ΔI$_c$ as a function of in-plane magnetic field applied at different angles with respect to the current direction and the corresponding color contour map for the same device. θ = 0° / ±180° (and ±90°) correspond to the in-plane magnetic field being perpendicular (and parallel) to the current direction. **d,** Dependence of ΔI$_c$ on temperature for $B_y$ = 12 mT for the same device. Inset shows a quadratic dependence $(T - T_J)^2$ of the ΔI$_c$ on temperature close to $T_J$. **e-f,** Calculation of the dependence of ΔI$_c$ on the angle of the in-plane magnetic field and the corresponding color map, performed from eq. (7) at $B_d$ = 22 mT and $B_c$ = 45 mT.

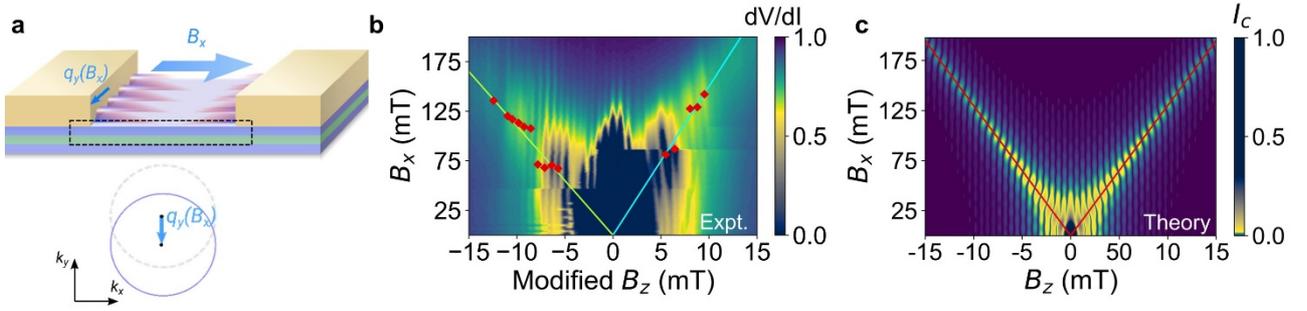

**Fig. 3| Observation of finite momentum pairing. a,** Schematic of the JJ in the presence of an in-plane magnetic field parallel to the current direction, leading to a shift of the Fermi surface in the *y*-direction. The waves illustrate the oscillations of the real part of the order parameter in the vicinity of the leads that occur due to the finite-momentum pairing. **b,** Dependence of $\frac{dV}{dI}$ on in-plane ($B_x$) and out-of-plane ($B_z$) magnetic field for a JJ device with $d$ = 350 nm showing the evolution of the interference pattern due to the applied in-plane magnetic field. The color corresponding to the normalized magnitude of $\frac{dV}{dI}$ is shown on the right side of each figure. The red diamonds correspond to the positions of the centers of peaks with respect to $B_x$ and the solid lines correspond to the linear fits used for determining the slopes of the side branches (see SI for details). **c,** Calculation of the evolution of the interference pattern expressed as a function of $B_z$, where color represents the value of $I_c$. The solid lines mark the side branches with the slope corresponding to the average of the experimentally obtained slopes $\left(\frac{\Delta B_x}{\Delta B_z}\right)_{avg} \approx 13$. We note that the qualitative behavior of $\frac{dV}{dI}$ is the same as that of $I_c$, including the periodicity with respect to $B_z$ and the slope of the side branches [30].

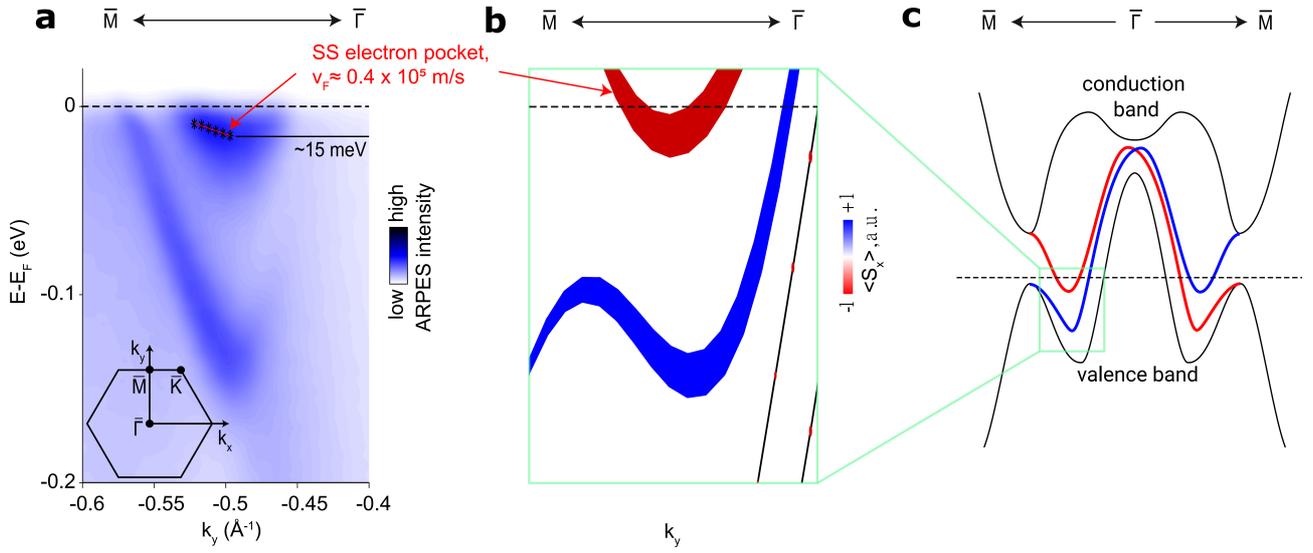

**Fig. 4| Spin-polarized surface states of NiTe$_2$. a,** ARPES dispersion ($\bar{\Gamma} - \bar{M}$ direction) shows two surface states crossing the Fermi level, one of them forming an electron pocket with a small Fermi velocity ($v_F$) and Fermi energy. Black stars indicate the maxima of the peaks extracted from energy distribution curves (EDCs), and the red line indicates the linear fit of the peak positions that was used to estimate the Fermi velocity for one of the surface states (see SI for more details). The spectra were measured with a photon energy of $h\nu$ = 23 eV and a horizontal linear polarization. The inset shows the surface Brillouin zone of NiTe$_2$ with the high-symmetry points noted. **b,** DFT calculation of the energy-momentum dispersion along the $\bar{\Gamma} - \bar{M}$ direction. Two surface states are highlighted, the line width is proportional to the the surface contribution and the component of the spin polarization perpendicular to the momentum is shown in color. **c,** Schematic picture showing the relative position of the surface states with respect to the conduction and valence bands.

18  He, J. J., Tanaka, Y. & Nagaosa, N. A Phenomenological Theory of Superconductor Diodes in Presence of Magnetochiral Anisotropy. *arXiv* **2106.03575** (2021).

19  Mukherjee, S. *et al.* Fermi-crossing Type-II Dirac fermions and topological surface states in $NiTe_2$. *Sci. Rep.* **10**, 12957, doi:10.1038/s41598-020-69926-8 (2020).

20  Ghosh, B. *et al.* Observation of bulk states and spin-polarized topological surface states in transition metal dichalcogenide Dirac semimetal candidate $NiTe_2$. *Phys. Rev. B* **100**, 195134, doi:10.1103/PhysRevB.100.195134 (2019).

21  Kodama, J. i., Itoh, M. & Hirai, H. Superconducting transition temperature versus thickness of Nb film on various substrates. *J. Appl. Phys.* **54**, 4050-4054 (1983).

22  Farrell, M. E. & Bishop, M. F. Proximity-induced superconducting transition temperature. *Phys. Rev. B* **40**, 10786-10795 (1989).

23  Tinkham, M. Introduction to Superconductivity. *Dover Publication Inc* (2004).

24  de Gennes, P. G. & Mauro, S. Excitation spectrum of superimposed normal and superconducting films. *Solid State Commun.* **3**, 381-384 (1965).

25  Clarke, J. The proximity effect between superconducting and normal thin films in Zero field. *J. Phys. Colloques* **29**, C2-3-C2-16 (1968).

26  Assouline, A. *et al.* Spin-Orbit induced phase-shift in $Bi_2Se_3$ Josephson junctions. *Nat. Commun.* **10**, 126 (2019).

27  Zhu, Z. *et al.* Discovery of segmented Fermi surface induced by Cooper pair momentum. *Science* **0**, eabf1077.

28  Yuan, N. F. Q. & Fu, L. Zeeman-induced gapless superconductivity with a partial Fermi surface. *Phys. Rev. B* **97**, 115139 (2018).

29  Hart, S. *et al.* Controlled finite momentum pairing and spatially varying order parameter in proximitized HgTe quantum wells. *Nat. Phys.* **13**, 87-93 (2017).

30  Chen, A. Q. *et al.* Finite momentum Cooper pairing in three-dimensional topological insulator Josephson junctions. *Nat. Commun.* **9**, 3478, doi:10.1038/s41467-018-05993-w (2018).

31  Gubin, A. I., Il'in, K. S., Vitusevich, S. A., Siegel, M. & Klein, N. Dependence of magnetic penetration depth on the thickness of superconducting Nb thin films. *Phys. Rev. B* **72**, 064503 (2005).

32  Clark, O. J. *et al.* Fermiology and Superconductivity of Topological Surface States in $PdTe_2$. *Phys. Rev. Lett.* **120**, 156401, doi:10.1103/PhysRevLett.120.156401 (2018).

33  Yuan, N. F. Q. & Fu, L. Topological metals and finite-momentum superconductors. *Proc. Natl. Acad. Sci. U.S.A.* **118**, e2019063118 (2021).
14